\documentclass[]{aa}  

\usepackage{graphicx}
\usepackage{txfonts}
\usepackage[colorlinks=true,linkcolor=blue,citecolor=blue]{hyperref}
\usepackage{url}

\begin{document}

   \title{Multiwavelength dissection of a massive heavily dust-obscured galaxy and its blue companion at z$\sim$2}

   \subtitle{}

   \author{M. Hamed\inst{1}
         \and
         L. Ciesla\inst{2}
         \and
         M. B\'ethermin\inst{2}
         \and
         K. Ma\l{}ek\inst{1,2}
         \and
         E. Daddi\inst{3}
         \and
         M. T. Sargent\inst{4}
         \and
         R. Gobat\inst{5}
          }

         \institute{National Centre for Nuclear Research, ul. Pasteura 7, 02-093 Warszawa, Poland
         \and
         Aix Marseille Univ. CNRS, CNES, LAM, Marseille, France
         \and
         CEA, Irfu, DAp, AIM, Universit\`e Paris-Saclay, Universit\`e de Paris, CNRS, F-91191 Gif-sur-Yvette, France
         \and
         Astronomy Centre, Department of Physics and Astronomy, University of Sussex, Brighton, BN1 9QH, UK
         \and
         Instituto de Fisica, Pontificia Universidad Cat\'olica de Valpara\'iso, Casilla 4059, Valparaiso, Chile}

   \date{Received 02 October 2020 / Accepted 15 January 2021}
 
  \abstract
  {}
  {In this work we study a system of two galaxies, namely Astarte and Adonis, at z $\sim$ 2 when the Universe was undergoing its peak of star formation activity. Astarte is a dusty star-forming galaxy at the massive-end of the main sequence (MS) and Adonis is a less-massive, bright in ultraviolet (UV), companion galaxy with an optical spectroscopic redshift. We analyze the physical properties of this system, and probe the gas mass of Astarte with its ALMA CO emission, to investigate whether this ultra-massive galaxy is quenching or not, and whether it has always been on the MS of star-forming galaxies.
 }
  {We use CIGALE - a spectral energy distribution modeling code - that relies on the energetic balance between the UV and the IR, to derive some of the key physical properties of Astarte and Adonis, mainly their star formation rates (SFRs), stellar masses, and dust luminosities. We inspect the variation of the physical parameters depending on the assumed dust attenuation law. We also estimate the molecular gas mass of Astarte from its CO emission, using different $\alpha_{CO}$ and transition ratios ($r_{31}$) and discuss the implication of the various assumptions on the gas mass derivation.}
  {We find that Astarte exhibits a MS-like star formation activity, while Adonis is undergoing a strong starburst (SB) phase. The molecular gas mass of Astarte is far below the gas fraction of typical star-forming galaxies at z=2. This low gas content and high SFR, result in a depletion time of $0.22\pm0.07$ Gyrs, which is slightly shorter than what is expected for a MS galaxy at this redshift. The CO luminosity relative to the total IR luminosity suggests a MS-like activity if we assume a galactic conversion factor and a low transition ratio. The SFR of Astarte is of the same order using different attenuation laws, unlike its stellar mass that increases using shallow attenuation laws ($\sim$ 1$\times$10$^{11}$ M$_{\odot}$ assuming a Calzetti relation versus $\sim$ 4$\times$10$^{11}$ M$_{\odot}$ assuming a shallow attenuation law). We discuss these properties and suggest that Astarte might be experiencing a recent decrease of star formation activity and is quenching through the MS following a SB epoch.
  }
   {}

   \keywords{galaxies: evolution - galaxies: high-redshift - galaxies: star formation - galaxies: starburst - infrared: galaxies - ISM:~dust, extinction.
               }

   \maketitle
%

\section{Introduction}

    Studying galaxy evolution throughout the cosmic time is a key element of modern astrophysics, and is crucial for our understanding of the life cycle of the progenitors of passive elliptical galaxies that we observe in the local Universe. Evidences suggest that the star formation rate (SFR) density has peaked around a redshift of $z \approx 2$ \citep[e.g.,][]{Hopkins06,Madau14,Bethermin2017,Gruppioni2020}, making this epoch (the cosmic noon) particularly interesting. Moreover, the cosmic noon epoch is when the dusty star-forming galaxies (DSFGs) \citep[e.g.,][]{Smail1997,Blain2002,Weiss2013,Casey2014,Donevski2020} contributed significantly to the star formation activity of the Universe \citep[e.g.,][]{Chapman2003, Chapman2005}. Furthermore, the dust-obscured star formation activity plays an important role at higher redshifts \citep[e.g.,][]{Takeuchi2005, Murphy2011, Bethermin2015, Bourne2017, Whitaker2017}. It is therefore crucial to study the massive DSFGs at higher redshift.\smallbreak 
    
    The affluence of multiwavelength data, especially the far infrared (FIR) detections from $Herschel$, played a central role in understanding how DSFGs evolve as a function of redshift. However, there are still controversies regarding how these galaxies build up their stellar masses. These controversies arise from the systematic uncertainties caused by the heavy dust attenuation in this type of objects \citep[e.g.,][]{Hainline2011, Michalowski2012}. This is caused by the sensitivity of the stellar mass estimate to the type of star formation history (SFH), the choice of the synthetic stellar population (SSP), and the assumed initial mass function (IMF). The debate about the accuracy of derived stellar masses of DSFGs was also discussed in details in \citet{Casey2014}.\smallbreak
        
    On the other hand, the growing number of ALMA observations in the recent years is providing an unparalleled help in constraining the evolution of DSFGs. These data are allowing us to build a comprehensive view of the role of these giant IR-bright sources by tracing their molecular gas and dust content \citep[e.g.,][]{Donevski2020}. The wealth of multiwavelength data also contributed to improve significantly estimating physical properties that govern such galaxies, by modeling their spectral energy distribution \citep[SED, e.g.,][]{Burgarella2005,daCunha2008,Noll2009,Conroy2013,Ciesla2014}. \smallbreak
    To build an SED, different aspects of a galaxy must be considered, most importantly the star formation history (SFH), the change of which has a strong impact on the derived SFR \citep[e.g.][]{Buat14,Ciesla17}, stellar populations of varied ages and metallicities, dust emission with different dust grain sizes and temperatures, nebular and synchrotron emissions, etc. Extinction caused by dust is critically important in any spectrum fitting of a galaxy, since it mutates the shape of the SED the most by absorbing a significant amount of the UV photons and thermally re-emitting them in the IR. This behavior can be modeled by assuming that dust absorbs the shorter wavelength spectrum of galaxies following attenuation laws, which are typically described by simple power laws with varying complexities, and is able to reproduce the observed extinction in galaxies of different redshifts and types. However, dust attenuation laws are not universal \citep[e.g.,][]{Wild2011,Buat2018,Kasia2018,Salim2020}. The need of different attenuation recipes is inevitable in order to reproduce the spectra of galaxies of different masses, IR luminosities and naturally the redshift. This makes it challenging to interpret some of the physical features especially when different attenuation laws can reproduce a good SED of a galaxy \citet{buat2019}. \smallbreak 
    A non-negligent fraction of galaxies exhibit a non-alignment and sometimes a total disconnection between the dust continuum and the stellar population \citep{Dunlop2017,Elbaz2018}. This directly challenges SED fitting techniques that rely on the energetic balance between the UV and the IR, since the key assumption for such techniques is that any physical property derived from one part of the spectrum should be valid for the entire galaxy. Several approaches were investigated to test the validity of this strategy, \citet{buat2019} suggested the decoupling of the stellar continuum from the IR emission by modeling their fluxes apart to compare the derived parameters such as the SFRs, dust luminosities and stellar masses with the ones derived using full SEDs. Statistical samples of such massive and dusty galaxies \citep[e.g.,][]{Dunlop2017, Elbaz2018, buat2019, Donevski2020} offer an important insight into the evolution of dust and gas mass through the cosmic time. However, the nature of these giants is not fully understood.\smallbreak
    
    The interstellar medium (ISM) is the most important element in understanding the physical processes of star formation itself, since it contains the building materials for future stars, most importantly the hydrogen. Hydrogen's density was found to be tightly correlated with the SFR, as suggested by \citet{Schmidt59} and investigated by \citet{Kennicutt98}. This correlation is known as the Schmidt-Kennicutt law, and it takes into account the gas in its molecular and atomic forms, albeit the molecular gas having the biggest impact. The mass of this gas can be estimated with the emission of the easily excited CO molecules \citep[e.g.,][]{Carilli2013,Weiss2013b,Decarli2019,Riechers2020}. Tracing the molecular gas with CO emission relies entirely on already established abundances in galaxies of the local Universe. Large interferometers such as ALMA offer unique opportunities to detect these emission lines with an unprecedented accuracy, the luminosity of which can give an estimate of the molecular hydrogen mass of a galaxy, typically using a conversion factor. On the other hand, conversion factors in high-redshift galaxies are highly debated (see \citealp{Bolatto13} for a comprehensive review).

    Nonetheless, an estimate of the molecular gas reservoir of galaxies at the high-mass end of the main sequence is crucial to characterise their star formation activity. For instance, \citet{Elbaz2018} showed that some galaxies exhibit a SB-like gas depletion time scale despite residing on the MS. \smallbreak
    
    Despite the growing number of detection of such heavily dust-obscured ultra-massive objects at high redshift, the progress of SED modeling, and the better comprehension of the high-redshift ISM, we still lack a full picture of how these galaxies form and quench. Were they always steadily forming stars along the MS? Or are they former SBs transiting to the red sequence through the main sequence?
    
    To answer these questions, it is essential to understand how is the star formation fueled by the gas in massive objects, and why does this activity cease. Quenching mechanisms are still not fully understood and they might be caused by AGN feedback or outflows \citep[e.g.][]{Cattaneo2009, Dubois2013,Combes2017} to environmental effects that can lead to gas stripping \citep[e.g.][]{Coil2008,Mendez2011}.\smallbreak
    
    In this paper, and motivated by the aforementioned questions, we analyze and interpret the multi-wavelengths observations of a pair of galaxies at z$\sim$2, with original COSMOS2015 catalog \citet{laigle16} IDs: 647980 and 648299, hereafter Astarte and Adonis. Astarte is an ultra-massive (M$_{\star} > 10^{11} M_{\odot}$), IR-bright galaxy, for which the CO emission is serendipitously detected with ALMA. Adonis is a low-mass galaxy bright in near-UV (NUV) and optical bands.\smallbreak
    The structure of this paper is as follows: in Section \ref{section2} we describe the data of the two galaxies analysed in this work. In Section \ref{section3.1} we probe the molecular gas of Astarte using its ALMA-detected CO emission line, and in Section \ref{section3.2} we investigate the morphology of this line compared to multiwavelength detection. In Section \ref{section3.3} we derive the physical properties of the two galaxies using SED fitting. the discussion and the conclusion are in Sections \ref{discussion} and \ref{conclusion} respectively.\newline
    Throughout this paper, we adopt the stellar initial mass function (IMF) of \citet{Chabrier03} and a $\Lambda$CDM cosmology parameters (WMAP7, \citealp{Komatsu2011}): H$_0$ = 70.4 km s$^{-1}$ Mpc$^{-1}$, $\Omega_{M}$ = 0.272, and $\Omega_{\Lambda}$ = 0.728.

\section{Observations}\label{section2}
    The system of Astarte and Adonis studied in this paper was initially part of a selection of z$\sim$2 galaxies at the high-mass end (M$_{\star}>10^{11}\,M_\odot$) of the main-sequence (MS) of star-forming galaxies \citep[e.g.,][]{Noeske2007,Daddi2010,Rodighiero2011,Schreiber15} detected by \emph{Herschel}/PACS observations of the COSMOS field (PEP survey, \citealp{Lutz2011}). In the COSMOS2011 catalog where the system was selected, the system is not deblended even in the optical and near-IR and appears as a single source. It is probably caused by the fact that this early catalog is mainly built using the $i$ band, where Astarte is particularly faint. The zCOSMOS survey \citet{Lilly2009} measured the spectroscopic redshift at the position of the HST/ACS source from zCOSMOS and found z$_{spec}=2.140$. In the more recent COSMOS catalog \citet{laigle16}, both z and near-infrared bands were used to detect and deblend the object. Adonis and Astarte have thus an individual measurements of their flux in the optical bands of Subaru, the NIR bands of VISTA, and the mid-IR with \emph{Spitzer}/IRAC.
    Astarte is detected at 250 and 350 $\mu$m with \emph{Herschel}/SPIRE using a 24$\,\mu$m prior \citet{Oliver2012}. The aforementioned deblending, coupled with the FIR detection of Astarte results in a low-mass low-SFR object (SFR = 37 $M_{\odot}$ $yr^{-1}$ with a stellar mass of 9.46$\times 10^{9}$ $M_{\odot}$), and a dust-obscured ultra-massive object (SFR = 131 $M_{\odot}$ $yr^{-1}$ with a stellar mass of 1.41$\times 10^{11}$ $M_{\odot}$), as estimated initially using LePhare \citet{Arnouts2011}. \smallbreak
   Astarte and Adonis were observed by ALMA as part of a program (2013.1.00914.S, PI: Bethermin) targeting a pilot sample of four massive z$\sim$2 main-sequence galaxies in band-7 continuum and their CO emission. The goal was to measure their gas and dust content and to compare their short-wavelength morphology with their CO and continuum morphologies.
   
   \subsection{NUV-IR observations}
    The NUV (rest-frame FUV) detections of our two galaxies are provided by the Canada France Hawaii Telescope (CFHT) in the \emph{u} band. Visible and NIR detections (rest-frame mid-UV to NUV) are obtained via the broad band Suprime-Cam of Subaru in the \emph{B, V, r, i$^+$} bands and the mid-IR data (rest-frame NIR) are from the IRAC camera of \emph{Spitzer}. The IR-bright Astarte has a MIPS detection at 24\,$\mu$m with a signal-to-noise ratio (hereafter S/N) > 20 and is very bright (S/N $\sim 20)$ in IR detections of \emph{Herschel} where the beam size is large. The 100\,$\mu$m observation from PACS does not detect any of the two galaxies, but provides an upper limit, which is taken into account by the SED fitting of Astarte since it constrains the far-IR part of the spectrum. 
    
\begin{figure}[t]
    \centering
    \includegraphics[width=0.5\textwidth]{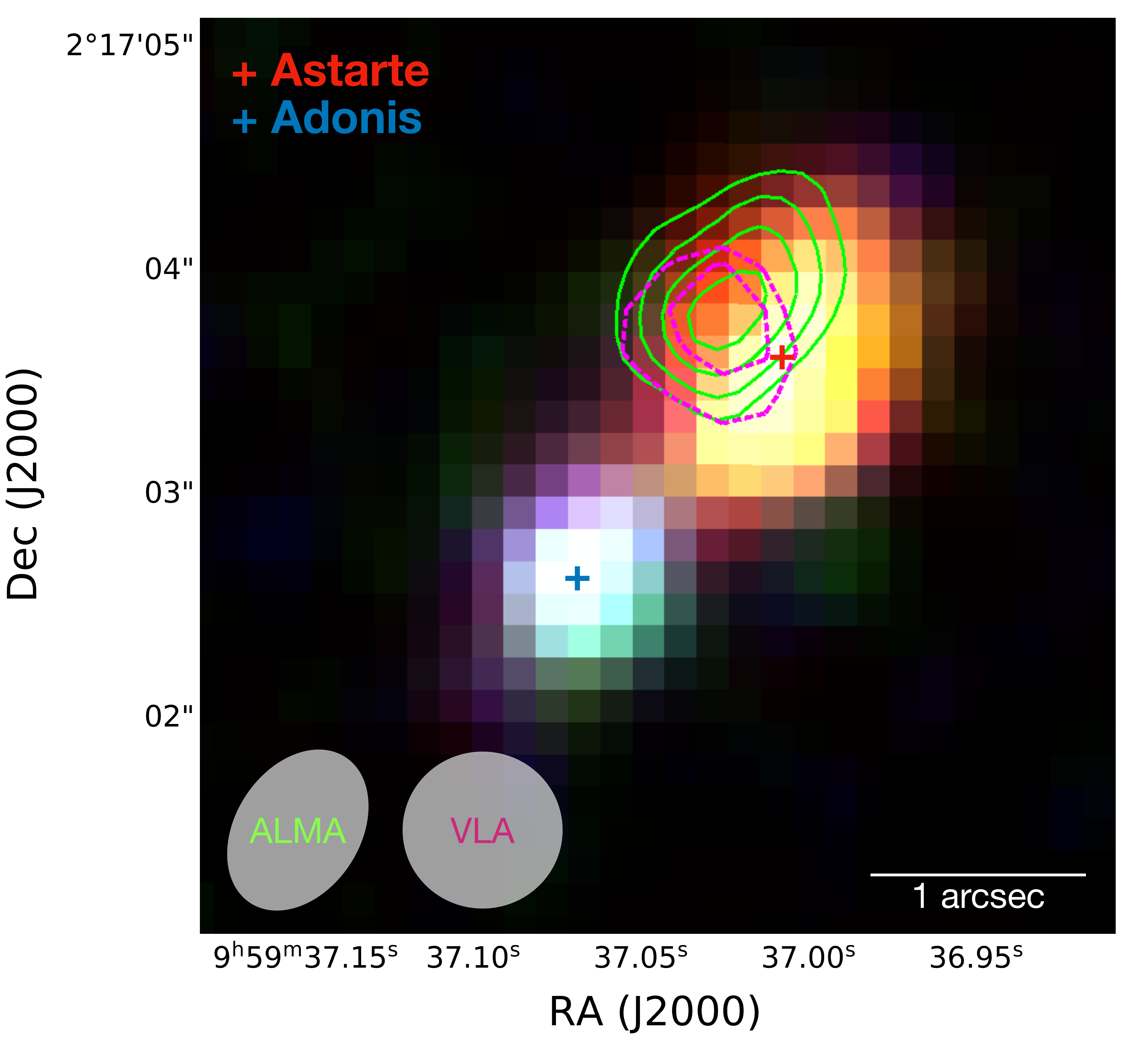}
    \caption{\footnotesize Integrated flux of ALMA-detected CO emission (green contours), along with the VLA-detected radio continuum at 3 GHz (magenta contours), on the RGB image (VISTA's Ks, H and J) of Astarte and Adonis. The beam size of ALMA is $0.78"\times0.50"$ (lower left beam). The beam FWHM of VLA is $0.75"$. The outermost contour of the CO integrated flux (green) is at 2$\sigma$ significance. The subsequent contours are in steps of 1$\sigma$ with the innermost contour showing 5$\sigma$. The magenta contours show 2 and 3$\sigma$ significance. The blue cross is centered on Adonis, the red cross is centered on Astarte.}
    \label{fig:Figure1}
\end{figure}

    The radio continuum of Astarte is tentatively detected with the Karl G. Jansky Very Large Array (VLA) in the S band at $\nu=$3\,GHz \citet{Smolcic2017}. This tentative detection was not included in the initial catalog of \citet{Smolcic2017}, since it falls just below their detection threshold of 5$\sigma$ (S/N=4.3). Adonis does not have any detection from VLA at 3 GHz. We thus estimated a 3$\sigma$ upper limit from the standard deviation in the cutout image around our two sources. The beam width of the VLA detection is $0.75\arcsec$, and the continuum is shown in Figure~\ref{fig:Figure1}.\smallbreak
    
    The \citet{Jin2018} catalog provides JCMT's fluxes at 850$\mu$m for both of our galaxies (2440$\pm$2519 $\mu$Jy for Astarte and 3910$\pm$2516 $\mu$Jy for Adonis). We refrain from using these super-deblended fluxes due to the high uncertainties, probably caused by the degeneracies in the deblending of such a close pair, and since the majority of flux is unexpectedly attributed to the smaller less IR-bright Adonis.
    Table \ref{tab:Table1} presents a summary of the available photometric data from different instruments of the two galaxies.

    \subsection{ALMA observation}
    Astarte was observed at 2.7\,mm with ALMA (band 3) with a time-on-source of 45 minutes using 32 antennas on September the $5^{th}$ of 2015, cycle-2 (P.I. M.B\'ethermin). We use the Common Astronomy Software Applications package and pipeline (CASA) v5.4\footnote{\url{https://casa.nrao.edu/}} \citep{McMullin2007} for flagging and to reduce the visibility data. The deconvolution was performed with the CLEAN algorithm using natural weighting for an optimal S/N. Multi-frequency synthesis mode of the line-free channels showed a non-significant continuum emission of the spectrum therefore its subtraction was not needed.
    In the deconvolution process, the cell size was set to $0.1\arcsec$. The achieved synthesized beam size is $0.78\arcsec \times 0.56\arcsec$, the velocity resolution of the cube is 21.36\,km\,s$^{-1}$ and the rms is 0.47\,mJy\,beam$^{-1}$\,km\,s$^{-1}$ per channel.

\begin{table}[!htbp]
  \begin{center}
  \tiny
    \begin{tabular}{l c c c c}
      \hline\hline
      &  &   &  Astarte & Adonis \\
      \hline
    COSMOS15 ID & & & 647980  & 648299 \\
    \hline
    redshift & & & z$_{phot} = 2.153$ & z$_{spec} = 2.140$\\
     \hline\hline
     Telescope/ & Filter & $\lambda$ &  $S_{\nu}$    & $S_{\nu}$\\
      Instrument&        & ($\mu$m)  &   ($\mu$Jy)     & ($\mu$Jy)      \\
      \hline\hline
     CFHT/      &  u     &  0.383    & $0.104\pm0.032$ & $0.492\pm0.032$\\
     MegaCam    &        &           &                 &                \\
     \hline
     Subaru/      & B &   0.446      &  $0.127\pm0.018$                 &   $0.596\pm0.032$   \\
     Suprime-Cam  &  V  &   0.548    &    $0.252\pm0.033$               &  $0.938\pm0.049$    \\
                  &  r & 0.629       &     $0.246\pm0.029$              &   $0.904\pm0.041$   \\
                  & i$^+$            &    0.768    &   $0.331\pm0.035$  &     $0.973\pm0.042$ \\
                  &  z$^{++}$        &    0.910    &   $0.719\pm0.062$  &    $1.329\pm0.063$  \\

     \hline
     VISTA/     & Y  &  1.02  &  $0.836\pm0.155$    & $1.519\pm0.162$  \\
       VIRCam   & J  & 1.25   & $2.691\pm0.175$     & $2.682\pm0.181$ \\
                & H  &  1.65  & $4.234\pm0.241$     & $3.243\pm0.254$ \\
                & Ks &  2.15  & $9.536\pm0.351 $    & $4.776\pm0.362$\\
     \hline
      \textit{Spitzer} & IRAC1 &  3.6           & $18.60\pm0.07$    & $3.70\pm0.10$ \\
                       & IRAC2 &  4.5           & $25.10\pm0.10$     & $2.80\pm0.13$\\
                       & IRAC3 &   5.8          & $25.10\pm2.00$     & $3.60\pm2.60$ \\
                       & IRAC4 &  8.0           & $15.30\pm3.30$     & -           \\
     \hline
     \textit{Spitzer} & MIPS1 &  24 &      \multicolumn{2}{c}{$351\pm17$}    \\
     \hline
     \textit{Herschel} & PACS & 100 &      \multicolumn{2}{c}{$<6734$} \\
     \hline
     \textit{Herschel} & SPIRE & 250 & $17792\pm744$ &  - \\
                 &       SPIRE & 350 & $16058\pm1026$&  - \\
                     
    \hline
  ALMA         &       band 3 & 3100 & \multicolumn{2}{c}{$<117$}\\
  \hline
     VLA             & S  & 1.3$\times10^5$ & $9.9\pm2.3$  & $<7.3$  \\
     \hline
    \end{tabular}
    \caption{\footnotesize Summary of the data of the two sources observed through the different instruments. $S_{\nu}$ is the flux in ($\mu$Jy). $\lambda$ is the center of the specific filter band.}
    \label{tab:Table1}
  \end{center}
\end{table}
\section{Results}
\subsection{Probing the molecular gas of Astarte}\label{section3.1}
In the data cube we find only one significant line and no significant continuum source in the field of view. The line extraction procedure along with the derivation of the luminosity and the gas mass are described in the following subsections.
\subsubsection{Line extraction}
\begin{figure}[h]
     \centering
        \includegraphics[width=0.5\textwidth]{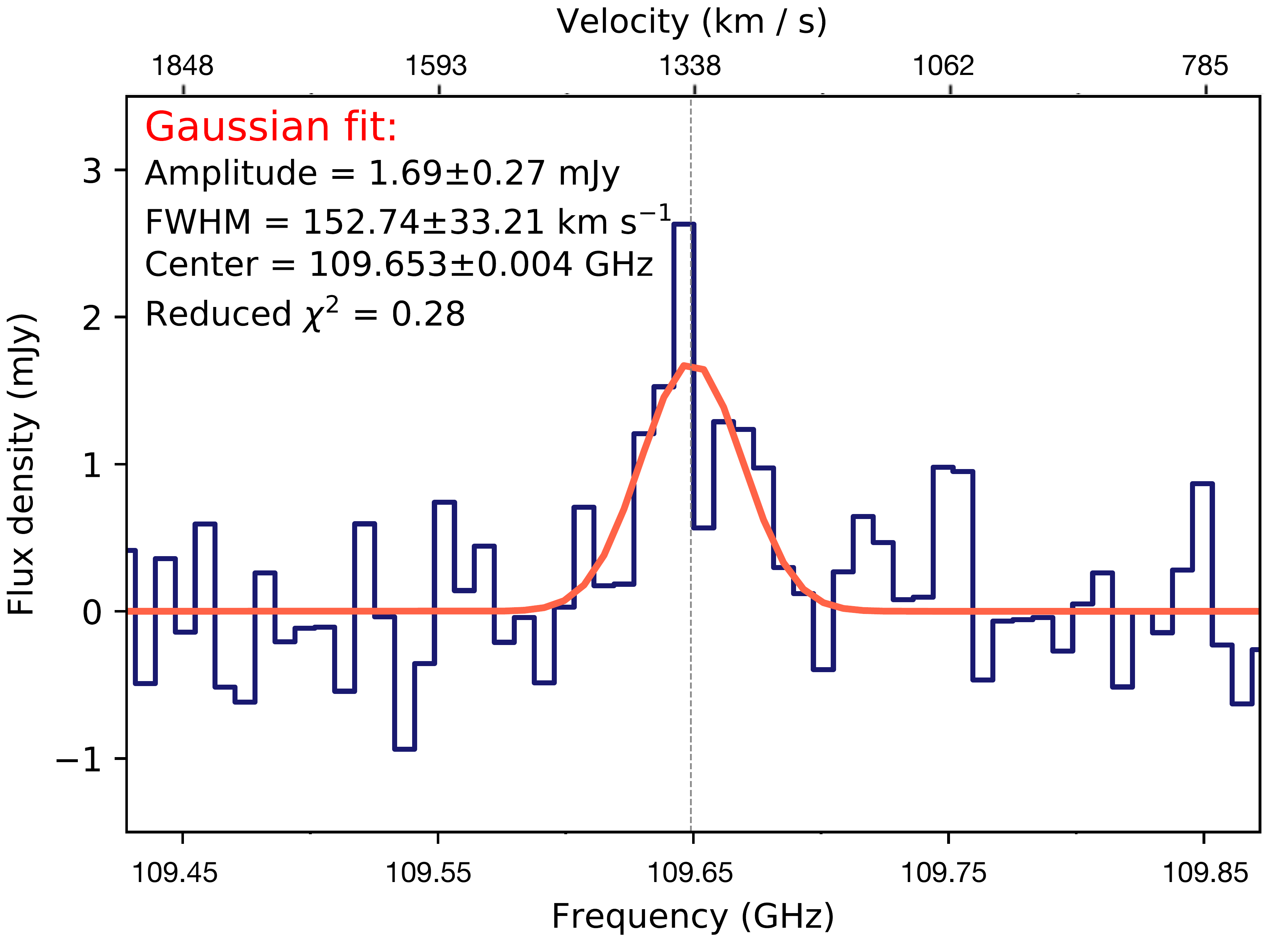}
        \caption{\footnotesize Spectral profile of Astarte with the redshifted CO(3$\rightarrow$2) line (dashed vertical grey line) and the Gaussian fit (red) with its properties.}
        \label{fig:Figure2}
    \end{figure}

    The ALMA-detected emission line of Astarte corresponds to the CO(3-2), with a peak at an observed frequency of $\nu_{obs} = 109.65$ GHz implying $z_{CO(3-2)} = 2.154$, which agrees with the photometric redshift $z_{phot, Astarte}=2.153^{+0.051}_{-0.058}$ \citet{laigle16}. This validates Astarte as the origin of the detected-CO emission.
    We do not detect Astarte in the continuum and measured a 3-$\sigma$ upper limit from the map of 0.117\,mJy. The expected flux densities from the SED modeling discussed in Sect.\ref{SED_results} are 0.007\,mJy and 0.049\,mJy for Adonis and Astarte, respectively. It is thus not surprising that none of our two sources are detected.
    The flux uncertainty was determined by deriving the standard deviation in the source-free pixels in the non primary-beam corrected map, since it has similar noise levels across the emission-free pixels (the noise at the central region is $\sim$ 2$\%$ higher than the outermost region of the map). The achieved S/N is 5.2 for the brightest channel of the CO(3-2) of Astarte.
    The emission line was extracted by fitting a Gaussian over the profile. The goodness of the Gaussian fit was verified with a $\chi^2$ test, its properties are summarized in Figure \ref{fig:Figure2} along with the redshifted CO(3-2) line. The full width at half maximum (FWHM) of the Gaussian is found to be 152.74$\pm$33.21 km s$^{-1}$.
    The spectroscopic redshift of the system at the position of HST detection found by by the zCOSMOS survey \citet{Lilly2009} is $z_{spec}=2.140$. This $z_{spec}$ corresponds to that of Adonis since only this UV-bright galaxy is detected with HST/ACS. Taking into account the redshift difference of Astarte and Adonis, the corresponding radial velocity difference $\Delta V$ is 1335$\, km\,s^{-1}$. This velocity difference is greater than what is found in interacting pairs of galaxies, which is typically $\Delta V < 350\, km\,s^{-1}$ \citep{Lambas2003, Alonso2004}. Outflows and absorption in the UV lines could account for few hundreds of km\,s$^{-1}$ \citet{Cassata2020}, or a division of the Hubble flow and peculiar motions, which could account for a significant velocity contribution if it is along the line of sight. Therefore, this does not eliminate a possible interaction between Astarte and Adonis.
    
\subsubsection{Line integrated flux and luminosity}
    The intensity is calculated by integrating over the Gaussian fit of the line, which is then converted to the apparent line luminosity (L$^{\prime}$), by using the expression from \cite{Solomon1997} that expresses $L^\prime$ with the integrated source brightness temperature in units of \text{$K\ km\ s^{-1}\ pc^2$}:
    \[ L^{\prime}_{line} = 3.25 \times 10^7 \times I \times\frac{D_L^2}{(1 + z)^3 \nu_{obs}^2}, \]
    where $D_L$ is the luminosity distance in (Mpc), $I$ is the intensity in (Jy km/s), and $\nu_{obs}$ is the observed frequency in (GHz). As a consistency test, we also estimated the integrated flux of the line using the moment-zero map, which was obtained by summing the channels where the emission line is detected. The line flux is measured in the moment-0 map using a 2-dimension Gaussian fit of the source. As shown in \citet{Bethermin2020}, there is no significant difference between this method and a fit in the uv plane for faint compact sources observed by ALMA. The resulting flux densities of the two methods are presented in Table\,\ref{tab:Table2}. There is 1.2\,$\sigma$ significant difference between the intensities derived by each methods. The spectrum is extracted at a single point assuming a point source, while the 2-dimension fit can recover the flux from an extended source. This small difference of flux suggests that our source could be marginally resolved. Hereafter, we use the flux from the moment-0 map, which takes this account. However, we cannot formally exclude another faint and diffuse component at larger scale considering the depth of our data. Figure \ref{fig:Figure1} shows the flux-integrated moment-0 map of Astarte represented by confidence levels contours. The size of the CO disk is $\sim74\ kpc$.
        \begin{table}[h]
      \begin{center}
        \begin{tabular}{ccc@{\extracolsep{\fill}}c}
          \hline\hline
           \small Peak flux & \small$I_{CO(3-2)}^{spec}$ & \small$I_{CO(3-2)}^{mom}$ & \small$L^{\prime}_{CO(3-2)}$ \\ 
        \small  density (mJy) & \multicolumn{2}{c}{(\small Jy km s$^{-1}$)} & \small (10$^9$ K km s$^{-1}$ pc$^2)$ \\ 
             \hline
            \small  $1.690\pm0.277$ & \small$0.251\pm0.062$  & \small$0.328\pm0.047$ &\small $8.508\pm1.219$ \\ 
             \hline
        \end{tabular}
        \caption{\footnotesize Summary of the CO(3-2) emission line properties of Astarte. $I_{CO(3-2)}^{spec}$ is achieved by integrating over the Gaussian of the emission line. $I_{CO(3-2)}^{mom}$ is the intensity derived from the moment-0 map.}
    \label{tab:Table2}
    \end{center}
   \end{table}
\subsubsection{Deriving the molecular gas mass}\label{alpha}

To derive the total mass of the molecular gas in a galaxy we assume that the H$_2$ mass is proportional to the CO(1-0) line luminosity which is the commonly used tracer of the cold star-forming molecular clouds, thanks to its small excitation potential requirement. The H$_2$ mass can be derived using a conversion factor $\alpha_{CO}$ \citep[e.g.][]{Downes2003, Greve2005, Tacconi2006, Carilli2013, Bothwell2013}:
\[M_{H_2} = \alpha_{CO}\ L^{\prime}_{CO(1-0)}\]
where M$_{H_2}$ is the mass of the molecular hydrogen in $M_\odot$, $\alpha_{CO}$ is the conversion factor and $L^{\prime}_{CO(1-0)}$ is the line luminosity in \text{$K\ km\ s^{-1}\ pc^2$}.
The practice of H$_2$ mass derivation with this method is very common especially for galaxies at high redshifts where information and spatial resolution is often limited. 
Our CO(3-2) line luminosity has to be converted to CO(1-0) luminosity using a luminosity line ratio $r_{31} = L^{\prime}_{CO(3-2)}/L^{\prime}_{CO(1-0)}$. We use $r_{31}=0.42\pm0.07$, which is the average ratio found for $z=1.5$ SFGs by \citealt{Daddi2015}. This results in: \[L^{\prime}_{CO(1-0)} = (2.03\pm0.59) \times 10^{10}\,K\,km\,s^{-1}\,pc^2.\]
To convert this luminosity into hydrogen mass, we use two conversion factors: $\alpha_{CO}=0.8$ and a galactic conversion factor of $\alpha_{CO}=4.36$. The first one manages to recover the molecular gas mass in starbursts (SBs) and submillimeter galaxies (SMGs), where the gas is efficiently heated by dust. The galactic conversion factor is suitable for normal main sequence galaxies \citep{DS98,Bolatto13,Carilli2013}. For $\alpha_{CO} = 0.8$, the mass of the molecular hydrogen is:
\[M_{H_2 (\alpha=0.8)} = (1.62\pm0.47) \times 10^{10}\ M_{\odot}.\]
Whereas $\alpha_{CO} = 4.36$ results in $M_{H_2} = (8.85\pm2.57)\times 10^{10}\ M_{\odot}$, five times larger gas reservoir than the one derived with $\alpha_{CO} = 0.8$. 

\subsubsection{Dynamical mass}
With the velocity FWHM of the CO(3-2) line, we use the method described in \citet{Bothwell13} to estimate the dynamical mass of Astarte. Assuming that a rotating disk is the origin of the detected line, the dynamical mass can be written as in \citet{Neri03}: \[M_{dyn}\ \ (M_{\odot}) = 4\times10^4\ \Delta V^2\ R / sin^2 (i),\] where $\Delta V$ is the FWHM of the line velocity, $i$ is the inclination angle of the disk and R is the radius of the disk in kpc.
For a random inclination of \(\langle i \rangle = 57.3^{\circ}\) \citep{Law2009}, the dynamical mass is found to be (1.11 $\pm${0.23}) $\times 10^{11}\,$M$_{\odot}$.\newline
The gas mass to dynamical mass ratio for a galactic conversion factor is therefore $M_{H_{2}}/M_{dyn}=0.76\pm0.33$. For $\alpha_{CO} = 0.8$, $M_{H_{2}}/M_{dyn}=0.15\pm0.07$.\smallbreak
The hydrogen mass derived assuming a galactic conversion factor is able to trace the dynamical mass of Astarte, despite the relatively low FWHM of the CO line, as it is for similar FWHM values in \citet{Bothwell2013}.

\subsubsection{CO emission morphology}\label{section3.2}
\begin{figure*}[t]
     \centering
    \includegraphics[width=1\textwidth]{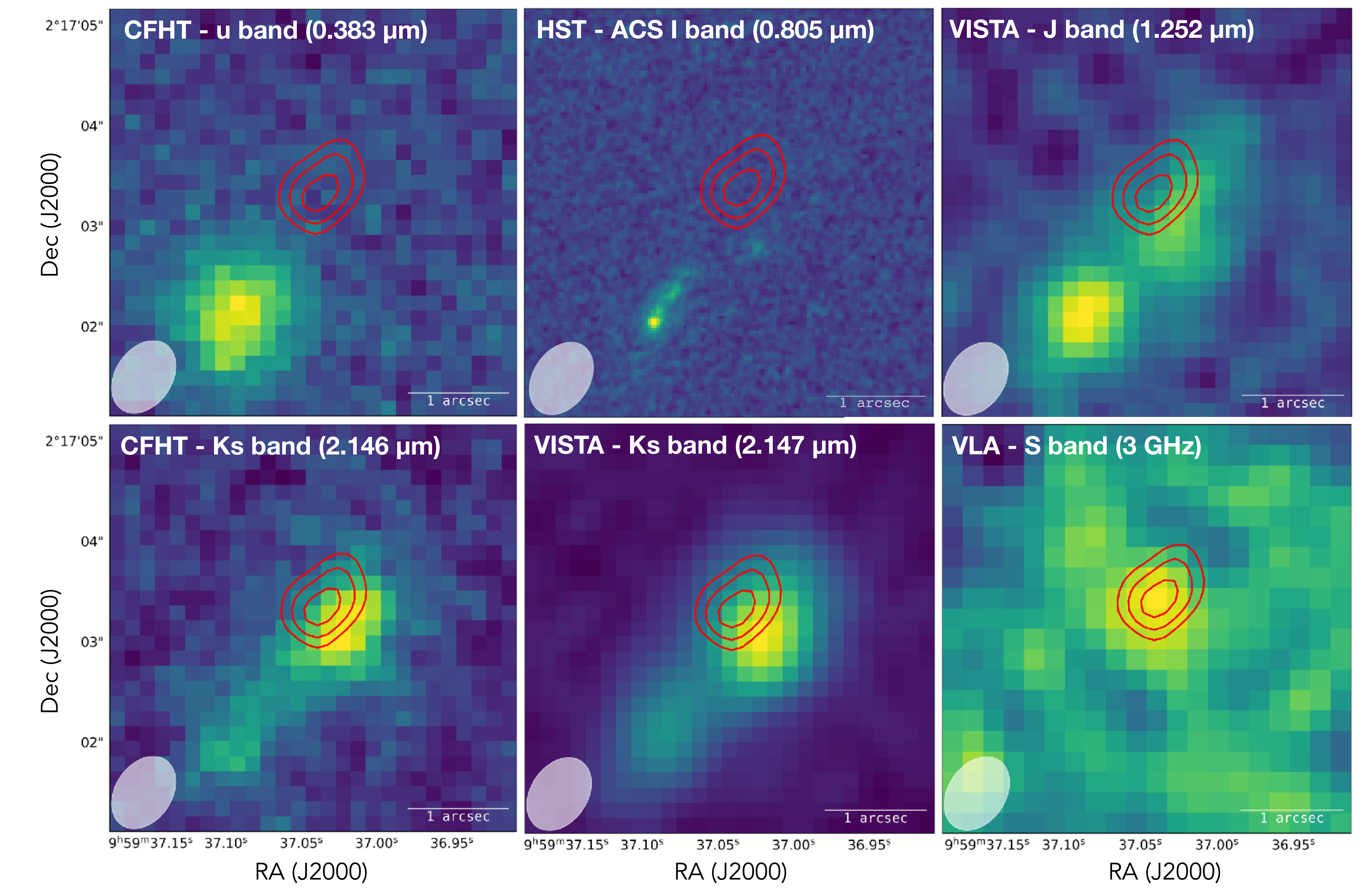}
    \caption{\footnotesize ALMA-detected CO(3-2) emission line contour map (red contours) of Astarte overlaid on detections from different telescopes/instruments at different bands as specified in every figure. From upper left to lower right: CFHT U band at 0.383 $\mu m$. HST's I band at 0.805 $\mu m$. VISTA J band at 1.252 $\mu m$. CFHT Ks band at 2.146 $\mu m$. VISTA Ks band at 2.147 $\mu m$ and on the VLA detection at 3GHz.  The outermost contour is 3$\sigma$, and the subsequent contours are in steps of 1$\sigma$ with red innermost contour showing 5$\sigma$. The beam size is 0.78"$\times$ 0.50". The white bar shows 1 arcsecond scale.
    }
    \label{fig:Figure3}
\end{figure*}

We investigate the morphology of the CO(3-2) emission line of Astarte in relation to other wavelength detections of the system, to closely study the association of the CO component with the UV, optical and IR components, as shown in Figure \ref{fig:Figure3}.\newline
HST's observation in the I band (mid-UV rest-frame) do not show Astarte due to its heavy dust obscuration. However, the young stellar population of the less-dusty Adonis is visible in HST's I band and is bright in the u band detection of CFHT (rest-frame far-UV) and in the J band of VISTA (rest-frame NUV). In the Ks bands of CFHT and VISTA, which correspond to rest-frame visible light, Adonis becomes less-bright, and is very faint at higher wavelengths observations of ALMA and VLA.\smallbreak
The dusty Astarte is not visible in the u band of CFHT (rest-frame FUV). It is however detected in the Ks bands of VISTA and CFHT showing a bright stellar population in the visible rest-frame wavelengths. A spatial offset (of $\sim$ 0.39") is visible between the ALMA-detected CO emission and the emission of the stellar population (observed in the Ks bands) of Astarte. \citet{Faisst2020} found an average offset between the COSMOS2015 catalogue and the Gaia reference frame of $\Delta (RA) = -63.9^{+70.7}_{-60.2}\, milliarcsec$ and $\Delta (Dec) = -1.4^{+80.4}_{-67.3}\, milliarcsec$. This systematic offset cannot explain the visible offset between the CO emission and the rest-frame optical counterparts of Astarte. Moreover, we show that the continuum detected by the VLA at 3 GHz of Astarte and its CO emission detected by ALMA are aligned, eliminating the possibility of a systematic error due to ALMA's synthesized beam size. \smallbreak

Although the original spectroscopic redshift of 2.140 (for both sources) found by zCOSMOS \citep{Lilly2009}, was derived from the visible range of HST's observation, where Astarte is not observed, ALMA offers a spectroscopic redshift for the latter (z$_{ALMA}$ = 2.154). This shows the importance of long-wavelengths detections especially for dust-obscured galaxies where the UV-NIR emission is heavily attenuated \citep{Schreiber18a, Wang2019}.

\begin{figure*}[t]
    \centering
    \includegraphics[width=1\textwidth]{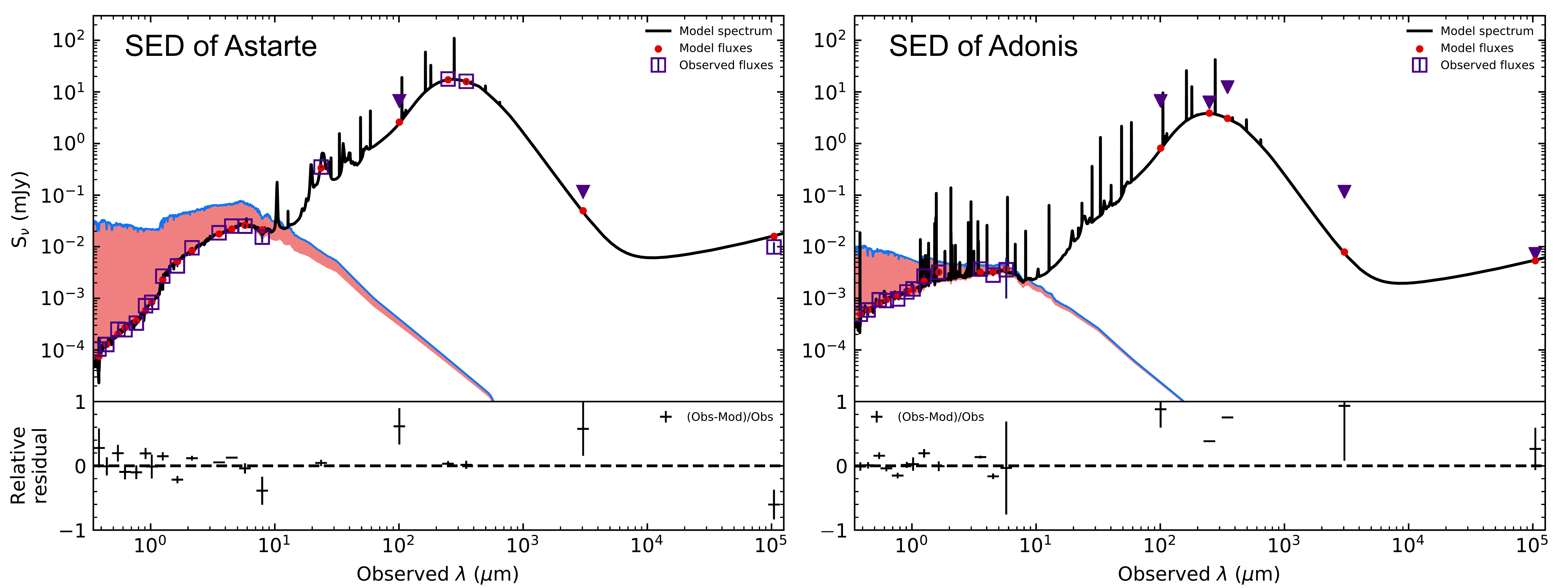}
    \caption{\footnotesize Best fits of the constructed SEDs of Astarte and Adonis along with their relative residuals. The SED of Astarte (left) is produced using LF17 attenuation law. The SED of Adonis (right) is produced using CF00 attenuation law. The best fit is shown in black. The unattenuated stellar emission is shown with the blue line. The filled region shows the difference between the unattenuated and the attenuated stellar emission, absorbed by dust. Red dots are the best fit values of the observations which are shown with the purple boxes. Upper limits are shown as purple triangles.
    }
    \label{fig:Figure4}
\end{figure*}
\subsection{SFRs, stellar masses and dust luminosities}\label{section3.3}

\begin{table*}[t]
  \begin{center}
 
    \begin{tabular}{l c c}
     \hline\hline
     Parameter & & Values \\
     \hline\hline
     \multicolumn{3}{c}{Star formation history \citet{Ciesla17}}   \\
     \hline
     Stellar age$^{(i)}$  & $age_{main}$ &  0.8 - 3.2\,Gyr by a bin of 0.2\,Gyr  \\
     e-folding time$^{(ii)}$ & $\tau_{main}$ &0.8, 1, 3, 5, 8\,Gyr \\
     Age of burst/quench episode & $t_{flex}$ & 5, 10, 50, 100\,Myr \\
     SFR ratio after/before  & $r_{SFR}$ & 10$^{-2}$, 10$^{-1}$, 0, 10$^1$, 10$^2$, 10$^3$ \\
    \hline\hline
    \multicolumn{3}{c}{Stellar synthesis population \citep{bc03}}   \\
    \hline
    Initial mass function & IMF & \citep{Chabrier03}\\
    Metallicity & $Z$ &  0.02 \\
    Separation age &    &  10\,Myr\\
    \hline\hline
    \multicolumn{3}{c}{Dust attenuation laws}   \\
    \hline
    \multicolumn{3}{c}{\citep{Calzetti00}}   \\
    \hline
    Colour excess of young stars & E(B-V) & 0.1 - 1 by a bin of 0.1 \\
    Reduction factor$^{(iii)}$ & f$_{att}$ & 0.3, 0.5, 0.8, 1.0\\
    \hline
    \multicolumn{3}{c}{\citep{CharlotFall00}, \textbf{\citep{lofaro17}}}    \\
    \hline
    V-band attenuation in the ISM & A$_V^{ISM}$ &  0.3 - 3 by a bin of 0.1 \\
    Av$_{V}^{ISM}$ / (A$_V^{BC}+A_V^{ISM})$ & $\mu$ & 0.3, 0.5, 0.8, 1 \\
    Power law slope of the ISM & &  -0.7, \textbf{-0.48} \\
    Power law slope of the BC &  & -0.7 \\
    \hline\hline
    \multicolumn{3}{c}{Dust emission model \citep{dl2014}}   \\
    \hline
    Mass fraction of PAH & q$_{PAH}$ &  1.77, 2.50, 3.19 \\
    Minimum radiation field & U$_{min}$ & 10, 25, 30, 40 \\
    Power law slope & $\alpha$ & 2 \\
    \hline\hline
    \multicolumn{3}{c}{Synchrotron emission}   \\
    \hline
    FIR/radio correlation coefficient &  &  2.3 - 2.9 by a bin of 0.1 \\
    Power law slope slope & $\alpha_{synchrotron}$ & 0.4 - 0.9 by a bin of 0.1 \\
    \hline
    \end{tabular}
    
    \caption{Input parameters used to fit the SEDs of Astarte and Adonis with CIGALE. (i) The stellar age is the age of the main stellar population. (ii) The e-folding time is the time required for the assembly of the majority of the stellar population. (iii) The reduction factor f$_{att}$ is the color excess in old stars relative to the young ones.}
    \label{tab:Table3}
    
    \end{center}
    \end{table*}

\begin{table}[t]
  \begin{center}
    \begin{tabular}{c c  c c c}
      \hline\hline
             &  Attenuation law  &   $\chi^2$ & reduced $\chi^2$ & BIC  \\
      \hline
                  & C00 & 43.22  & 2.12 & 73.18 \\
         Astarte & CF00 & 18.34  & 0.97 & 51.30 \\
                  & LF17 & 16.06 & 0.84 & 49.01 \\
      \hline
                  & C00 & 16.28 & 1.10 & 46.78\\
         Adonis & CF00 & 10.51 & 0.70 & 41.01 \\
                  & LF17 & 13.99 & 0.93 & 44.49 \\
         \hline
    \end{tabular}
    \caption{\footnotesize Comparison between the quality of fits of Astarte and Adonis produced with CIGALE with the three attenuation laws.}
    \label{tab:Table4}
  \end{center}
\end{table}

\subsubsection{The SED modeling}
We use the SED modeling code CIGALE\footnote{\url{http://cigale.lam.fr}} \citep{Boquien19} to derive the physical properties of our sources.
The code allows to model galaxies' SED from the UV to the radio wavelengths, taking into account the energetic balance between the emission absorbed by dust in the UV-visible range and the IR emission. CIGALE offers a variety of modules for each physical process a galaxy may undergo. The modules that we use in our SED fitting procedures are described below.

\subsubsection{Stellar component}
To model the stellar component of Astarte and Adonis, we use the stellar population synthesis of \citet{bc03}. This stellar library computes the direct stellar contribution to the spectrum (UV-NIR range) by populating the galaxy with young and old stars of different masses, as well as the required gas mass that will produce such population. This model was developed based on observations of nearby stellar populations and it describes well the various stellar emissions that one expects to encounter in any galaxy. These models depend on the metallicity and the separation age\footnote{Age of the separation between the young stellar population and the old one.}. We use a solar-like metallicity and we take into account nebular emission since they contribute to the total SED model from the UV to NIR.\smallbreak
Different stellar demographics must be modeled with an appropriate SFH in any SED modeling, since it is critical to estimate the contribution of the young and old stars to the total flux. An appropriate SFH is key to derive the SFR of a galaxy as it strongly depends on the assumptions made \citep{Ciesla17}. CIGALE offers different SFH scenarios varying from the simple delayed SFH to more complex ones containing episodes of bursts or sudden drops in SFRs. We use the SFH proposed by \citet{Ciesla17} which is a combination of a smooth delayed buildup of the stellar population to model the long term SFH of a galaxy, and a recent flexibility in the last few hundred Myrs to allow for recent and drastic SFR variations (burst or quench). This SFH model has been proven to limit biases by decoupling the estimations of the stellar mass, mainly constrained by rest frame NIR data, from the SFR, constrained by UV and IR data \citep[e.g.,][]{Ciesla16,Ciesla18,Schreiber18b,Schreiber18a}.
This kind of SFH was used in the study of high-redshift ($z<3$) passive galaxies to model their SED \citep[e.g.,][]{Schreiber18b,Schreiber18a,Merlin18}. We limit the recent burst/quench episodes to the last 100 Myrs of the life of our sources. The recent burst is motivated by the ALMA detection, however, it is important to note that this burst makes it difficult to constrain the past SFH. The burst part of the SFH is usually responsible for fitting the UV data, whereas the previous SFH (delayed) is driven by for the older stellar population, manifested in the visible part of the SED.

\subsubsection{Attenuation laws}

Two prominent attenuation laws are the ones of \citet{Calzetti00} (hereafter C00) and \citet{CharlotFall00} (hereafter CF00). They are widely used in the literature, and along with their alternations can describe the behavior of the extinction in the UV to NIR caused by dust.\newline C00 and its recipes is in its core equivalent to reducing the short-wavelength flux coming from a stellar population by an opaque screen, with the opacity being dependant on the total extinction of the stellar emission at the B and V bands.\newline Another approach is CF00 power-law which is fundamentally different from C00: it attributes different attenuation to the ISM and to the birth clouds (hereafter BC). This makes the dust more effective at absorbing the UV light since the young stellar emission has to pass through the dust in the BC and the ISM. Stars that are older are attenuated only by the ISM dust.
The CF00 approach is slightly more complex and physical than C00 for high redshift ultra dusty galaxies, embodying different dust distributions and densities throughout a galaxy. 

C00 and CF00 rely for their efficiency of attenuating the stellar population on power-laws for their slopes. The power-law slopes for BCs and ISM in CF00 were originally fixed at -0.7 each. \citet{lofaro17}'s recipe (hereafter LF17) of CF00 was tuned by assuming a power-law for the slope of the attenuation in the ISM equal to -0.48. This recipe provides a steeper attenuation curve at shorter wavelengths. In this work we use these three attenuation laws and compare their best fits and their effects on deriving the physical properties of our sources.\smallbreak
To assess which attenuation laws to use when different modules can produce good and comparable fits, we employ the Bayesian information criterion (BIC), defined as the $\chi^2 + k\ln{n}$, where $\chi^2$ is the non reduced goodness of the fit, $k$ is the degrees of freedom of the model and $n$ is the total number of photometric fluxes used in the fit of the galaxy. We then evaluate the preference of a model over the other one by calculating the difference between their BICs: $\Delta BIC > 2$ translates into a notable difference between the two laws and the fit with the lowest $\chi^2$ is preferred. This method was used by \citet{Ciesla18} for carefully choosing successful scenarios of SFHs of quenching galaxies, and by \citet{buat2019} for assessing SEDs of $z\sim2$ ALMA-detected galaxies.

\subsubsection{Dust emission}

To model the dust emission we use the \citet{dl2014} IR emission models, which was calibrated using high resolution observation of the Andromeda galaxy. \citet{dl2014} considers a variety of dust grains heated by different intensities coming from the stars and the photodissociation regions, and is an improved version of the older \citet{dl2007} model by varying dust opacity across the radius of a galaxy. This IR model was successful in reproducing dust emissions of millions of \emph{Herschel}-detected galaxies as a part of the HELP project \citep{Kasia2018}.

\subsubsection{Synchrotron emission}

The VLA detection at 3\,GHz of Astarte and Adonis allow us to model the synchrotron emission of our objects, taking into account a non-thermal power law of the synchrotron spectrum and the ratio of the FIR/radio correlation.
The different parameters used to build our SEDs are shown in Table \ref{tab:Table3}.

\subsubsection{SED fitting results}\label{SED_results}
In the case of Astarte, CF00 and LF17 attenuation laws result in best SED fits over C00. The BIC of every model was calculated and is shown in Table \ref{tab:Table4} along with the other quality of fit assessments. $\Delta BIC_{(C00, CF00)}$ = 21.88 and $\Delta BIC_{(CF00, LF17)}$ = 2.29, this privileged the best fit produced with LF17 attenuation law and therefore was taken into account in deriving the physical properties. Despite the uncertainties on any assumed SFH model, the adopted SFH here fitted the short wavelength data the best. \smallbreak
For the less-massive Adonis, CF00 gave overall better fits than C00 and LF17, with $\Delta BIC_{(LF17, CF00)}$ = 3.48 and $\Delta BIC_{(C00, CF00)}$ = 5.77. The best SED of Astarte and Adonis are shown in Figure \ref{fig:Figure4}. The signature of dust attenuation is clear in the two SEDs, where the heavily dust-obscured Astarte has more attenuation of its overall stellar mass than it is the case with Adonis. The derived properties of both galaxies are shown in Table \ref{tab:Table5}. The L$_{IR}$ of Astarte of the order of 10$^{12}$, qualifies it to be an Ultra Luminous IR Galaxy (hereafter ULIRG). While Adonis's poor dust content is manifested in the weaker IR luminosity and lower dust mass.  
\smallbreak

To closely inspect the visible dissociation of the gas and the stellar population in Astarte, we follow the method used in \citet{buat2019} by dissecting the stellar continuum and the IR emission apart and comparing their derived properties with the ones obtained using full SEDs. Taking into account the UV-NIR data (0.3 - 8 $\mu m$), the best fit for the stellar continuum was obtained with the C00 law, with $\Delta BIC_{(CF00,C00)} = 8.3$ and $\Delta BIC_{(LF17,C00)} = 2.7$. The better quality fit of the stellar continuum produced using C00 is expected, since this power-law effectively attenuates the young stellar population, while the other two laws can be equally efficient in attenuating the older stars, a behavior that translates into a rise in the near IR absorbed light and therefore a rise in the total IR emission.
The IR luminosity derived from the stellar continuum gives $L_{dust}=(2.43\pm1.01)\times 10^{12}\ L_{\odot}$, relatively close to $L_{dust}$ derived from the full SED. The stellar mass derived from the stellar emission gives $(1.3\pm0.2)\times10^{11}\ M_{\odot}$ and the SFR$_{(UV-NIR)}=430\ M_{\odot}yr^{-1}$. From the IR data (MIPS - ALMA continua), we get $L_{dust}=(3.25\pm0.08)\times 10^{12}\ L_{\odot}$, consistent with the one derived with the full SED. This result is in agreement with the results of \citet{buat2019} where they found consistent dust luminosities derived from both the full SED and the IR data, while $L_{dust}$ deduced from the stellar continuum was underestimated.

\begin{table}[t]
  \begin{center}
    \begin{tabular}{l c c}
      \hline\hline
      Physical property &  Astarte & Adonis\\
      \hline
      redshift   & $z_{CO}=2.154$  & $z_{spec}=2.140$ \\
      L$_{IR}\ (10^{12}\ L_{\odot})$ & $3.16\pm 0.06$ & $0.62\pm0.04$  \\
      SFR$\ (M_{\odot}/yr)$ & $395\pm20$ & $129\pm59$\\
      M$_{\star} (M_{\odot})$& $(3.74\pm0.19)\times10^{11}$ & $(9.37\pm1.76)\times 10^{9}$\\
      M$_{dust} (10^{9} M_{\odot})$ & $1.01\pm0.11$ & $0.86\pm0.13$ \\
         \hline
    \end{tabular}
    \caption{\footnotesize Summary of the physical properties obtained for Astarte and Adonis obtained with CIGALE.}
    \label{tab:Table5}
  \end{center}
\end{table}

\section{Discussion}\label{discussion}

\begin{figure}[t]
    \centering
    \includegraphics[width=0.5\textwidth]{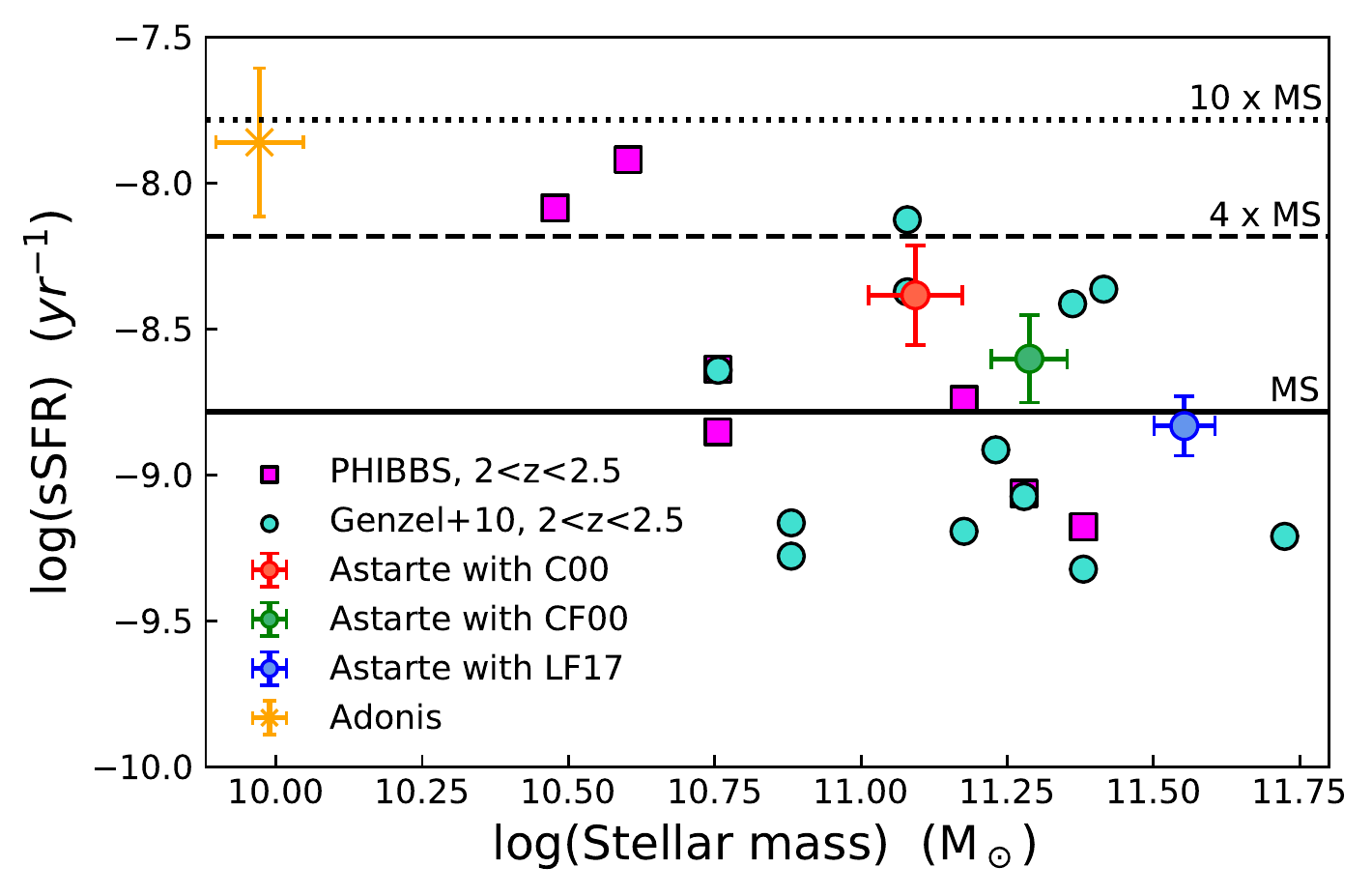}
    \caption{\footnotesize Relative position of the sSFR (SFR/M$_{\star}$) and stellar mass of Astarte using the three attenuation laws to the main sequence of \citet{Schreiber15} at z=2.  The yellow star shows the relative position of Adonis to the MS. Magenta squares denote PHIBSS CO-detected SFGs at z$\approx2$ \citep{tacconi13}. Turquoise circles are ULIRGs at 2<z<2.5 from \citet{Genzel10}. The solid line shows the MS of \citet{Schreiber15}. The dashed and dotted lines are $MS \times 4$ and $MS \times 10$ respectively.}
    \label{fig:Figure5}
\end{figure} 
\begin{figure}[t]
     \centering
    \includegraphics[width=0.5\textwidth]{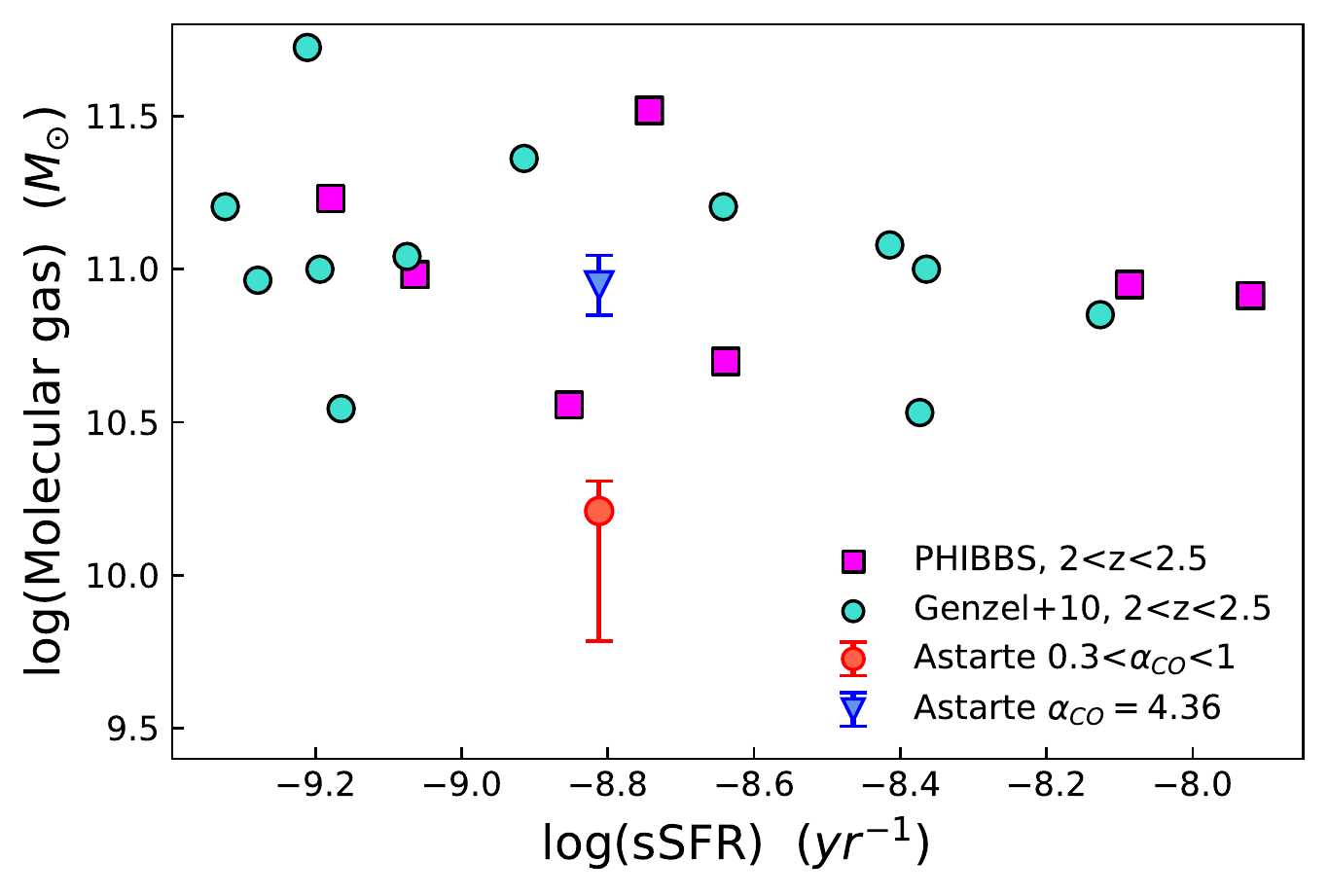}
    \caption{\footnotesize Molecular gas masses derived with the CO conversion factor versus the sSFR. The magenta squares are from T13. Turquoise circles are ULIRGs at 2<z<2.5 from G10. The red circle shows the position of Astarte with the used $\alpha_{CO}=0.8$, and the associated error bar shows the variation of the molecular gas mass using 0.3<$\alpha_{CO}$<1. The blue triangle shows the molecular gas of Astarte for a galactic CO conversion factor of 4.36.}
    \label{fig:Figure6}
\end{figure}

\begin{figure}[t]
     \centering
    \includegraphics[width=0.5\textwidth]{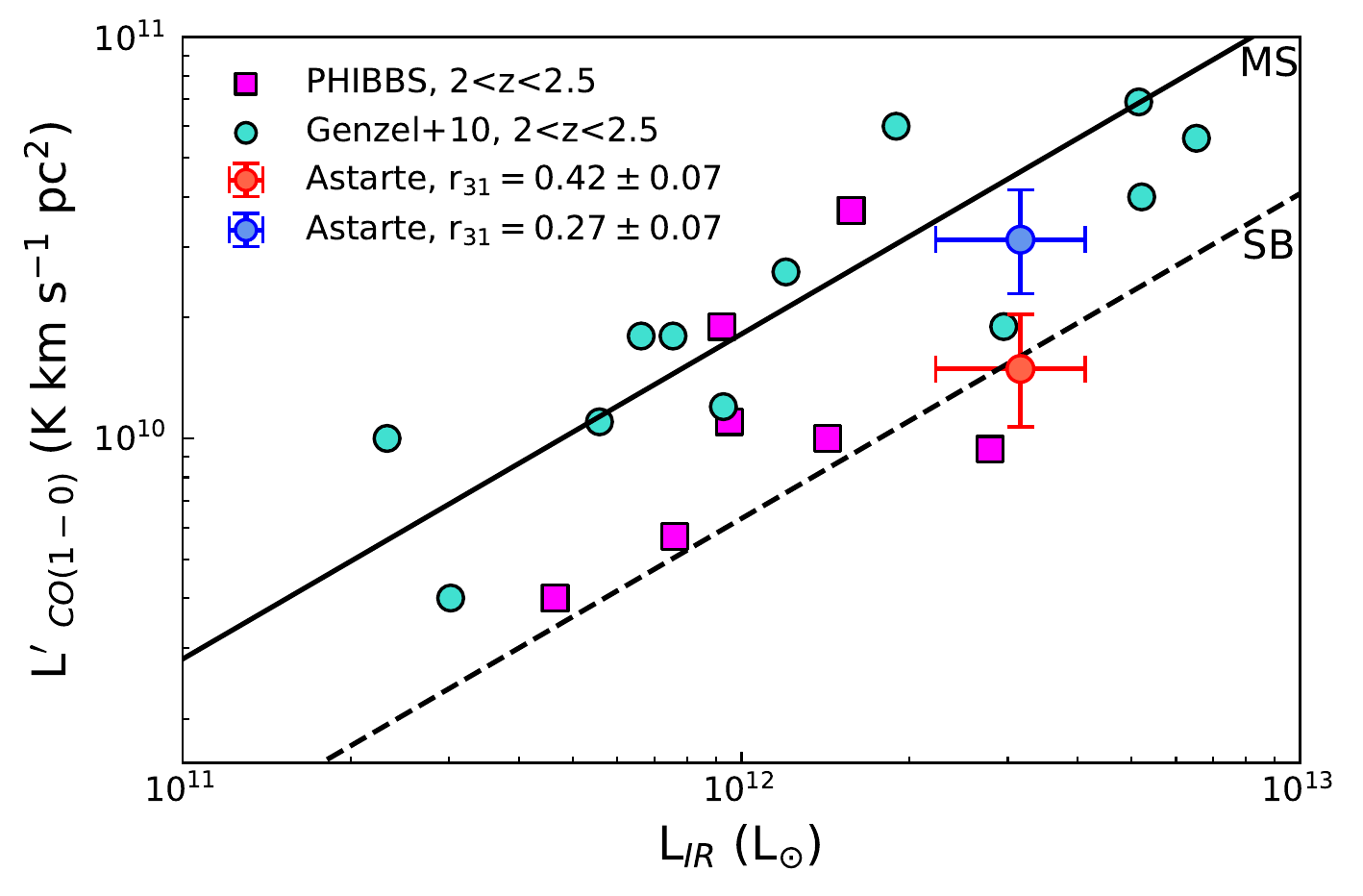}
    \caption{\footnotesize Correlation between CO(1-0) luminosities and the total IR luminosity. The magenta-filled squares are from T13 and turquoise-filled circles are the sources from G10. The solid black line is the linear regression for MS galaxies \citep{Sargent14} and the dashed line is that for SBs (the regression lines are from the complete sample in \citealt{Sargent14}).
    }
    \label{fig:Figure7}
\end{figure}

Figure \ref{fig:Figure5} shows the relative position of our galaxies to the MS of \citet{Schreiber15}. Adonis lies on 10$\times$MS qualifying it to be a strong starburst, despite its relatively low SFR. While being a starburst, this type of source cannot be detected even by the deepest 3\,mm ALMA survey (expected flux of 7\,$\mu$Jy), which have a 1-$\sigma$ noise of 9.7\,$\mu$Jy \citet{Gonzales2020}.\newline 
Astarte is a MS galaxy with all the different attenuation recipes used. However, there is a clear difference concerning the position of Astarte relative to the MS as a result of the three attenuation laws. This is attributed to the significant difference in the derived stellar masses, with CF00 and LF17 attenuation laws resulting in larger stellar mass than C00 due to the highest attenuation in NIR. This contributes to a lower specific SFR (sSFR = SFR/M$_{\star}$) since SFRs do not differ significantly with the three laws.\smallbreak

Host halo masses of such z$\sim$2 \emph{Herschel}-detected massive MS galaxies were investigated in \citet{Bethermin2014} using clustering and X-ray stacking and were found to reside in halos of $>$10$^{13}$ M$_\odot$. 
Such halo masses are also expected from the stellar mass to halo mass relation \citep{Behroozi13, Durkalec18, Behroozi19}. Astarte is $\sim$4 times less massive than the average central galaxies at z$\sim$1 of \citet{Hilton2013} and \citet{Burg2013}, which indicates that such MS giant galaxies continue to grow either through in situ star formation or accretion of other galaxies through the cosmic time until lower redshifts.\smallbreak

We compare Astarte with CO-detected samples of the same redshift range from \citet{Genzel10} (labeled G10) and \citet{tacconi13} (labeled T13). These samples of SFGs have constrained CO detections, and well-investigated physical parameters (in \citealp{Sargent14}).\smallbreak
We compare the molecular gas mass of Astarte derived from the CO emission line with that of G10 and T13 galaxies in Figure~\ref{fig:Figure6}. G10 uses a galactic conversion factor for SFGs and $\alpha_{CO}=1$ for ULIRGs, while T12 adopts a galactic conversion factor for all their sources. Our choice of $\alpha_{CO}=0.8$ underestimates the molecular gas mass of Astarte than the galaxies of G10 and T13. However, $\alpha_{CO}=4.36$ produces a higher molecular gas mass with respect to its sSFR.\smallbreak 

In Figure~\ref{fig:Figure7} we show the correlation between CO luminosities and the total infrared luminosities. IR luminosities were derived from the SFRs of all the sources (G10, T13 and Astarte) using the Kennicutt relation \citep{Kennicutt98}. The initial choice of $r_{31} = 0.42\pm0.07$ (the average in \citealp{Daddi2015}) places Astarte on the SB line from \citet{Sargent14}, contradicting its SED result. We therefore investigate the lowest excitation ratio from \citet{Daddi2015} of $r_{31} = 0.27\pm0.07$. This lower ratio moves Astarte closer to the MS within the error bars.\smallbreak
Using the total molecular gas mass of Astarte derived with the least excited CO(3-2) from \citet{Daddi2015} ($r_{31} = 0.27\pm0.07$), and assuming a galactic conversion factor, we estimate the gas fraction $f_{gas} = M_{gas} / (M_{gas}+M_{\star})$ = $(0.27\pm0.07)$. Although this falls within the lower limits of typical molecular gas fractions found in SFGs at z$\approx$2 in \citet{Santini2014} and \citet{Bethermin2015}, an $\alpha_{CO}$ adapted for SB with a higher r$_{31}$ ratio reduces the gas fraction significantly.\smallbreak 
The gas mass derived with a galactic conversion factor gives Astarte a rather small depletion time of $0.22\pm0.07$ Gyrs, making it very efficient at forming stars (for comparison, SBs have a depletion time of the order of $\sim100$ Myrs, (see Fig. 10 in \citealt{Bethermin2015}).
Recently, \citet{Elbaz2018} found that compact SFGs on the MS with a relatively low depletion time are not uncommon. These active ultra-massive objects can be hidden at the higher end of the tail of the MS. The average depletion time for \citet{Elbaz2018} galaxies is around 0.25 Gyrs, and although we do not detect the continuum of Astarte with ALMA, its CO emission is compact as it is the case for the continuum of ALMA-detected galaxies from \citet{Elbaz2018}. This is also confirmed in \citet{Puglisi2019} where compact massive galaxies at the top of the MS exhibit high SFRs at their cores following their SB epoch.\smallbreak

\section{Conclusion}\label{conclusion}
In this paper, we analyzed two galaxies, Astarte and Adonis, at the peak of the SFR density using multiwavelength dissection combining ALMA observations with UV-submm SED modeling. We investigated the molecular gas content of Astarte through the ALMA detection of its CO(3-2) emission, relying on different excitation ratios of $L^{\prime}_{CO(3-2)}/L^{\prime}_{CO(1-0)}$ and different CO conversion factors. A galactic conversion factor when used along with the least excitation ratio from \citet{Daddi2015} confirmed the relative position of Astarte to the MS, as found from its SED modeling. Although the obtained gas fraction is on the lower limits of that in MS galaxies \citep{Santini2014, Bethermin2015}, a possible explanation might be that the CO(3-2) instantaneous emission does not fully recover the molecular mass and the dynamics of Astarte, due to its weak excitation \citep{Daddi2015}. Detections of other transition levels of CO would be helpful to better constrain the molecular mass of Astarte, and therefore its physical characteristics.\smallbreak

The physical dissociation of the CO line and the rest-frame stellar population in Astarte was also investigated as done in \citet{buat2019}, by deriving physical properties from the stellar emission (UV-NIR) and the IR emission apart. As in \citet{buat2019}, the dust luminosity derived from the full SED is in agreement with the one derived from the the IR emission, while $L_{dust}$ derived from the stellar emission is slightly underestimated. Furthermore, C00 attenuation law was preferred when fitting the stellar continuum only. This is consistent with the results of \citet{buat2019} for galaxies with the same radii extension of rest-frame stellar emission and ALMA-detected emission. 
LF17 attenuation law, which was tuned for ULIRGs at $z\sim2$ succeeds the most in mimicking the dust attenuation of Astarte when fitting the whole UV-submm SED, this however results in higher stellar mass.

The molecular mass of Astarte, obtained with a galactic conversion factor and the lowest excitation ratio from \citet{Daddi2015}, contributes to $\sim0.9$ of its total dynamical mass, which is a larger contribution than what is found in ULIRGs of \citet{Neri03}. SFRs and stellar masses derived from the SED fittings show that Adonis is a SB galaxy, while Astarte is on the MS of SFGs. However, the small gas fraction makes Astarte very efficient in forming stars, whose depletion time is an order of magnitude lower than what is expected in typical MS galaxies \citep{Bethermin2015}. This SB-like star formation activity on the MS was found for massive compact SFGs in \citet{Puglisi2019}, in their post-SB phase. Low depletion times of MS massive galaxies were also found in \citet{Elbaz2018}, confirming that Astarte is caught in the middle of quenching following an earlier SB activity.\smallbreak 

Central galaxies at z$\sim$1 from \citet{Hilton2013} and \citet{Burg2013} are $\sim$ 4 times more massive than Astarte. This indicates that even massive objects that are on the high-end of the MS, when the Universe was undergoing its peak in the star formation rate density, continue their mass assembly down to lower redshift.

\begin{acknowledgements}
 M.H. and K.M have been supported by the National Science Centre (UMO-2018/30/E/ST9/00082). We acknowledge and thank the referee for a thorough and constructive report, which helped improving this work. This paper makes use of the following ALMA data: ADS/JAO.ALMA\#2013.1.00914.S. ALMA is a partnership of ESO (representing its member states), NSF (USA) and NINS (Japan), together with NRC (Canada), MOST and ASIAA (Taiwan), and KASI (Republic of Korea), in cooperation with the Republic of Chile. The Joint ALMA Observatory is operated by ESO, AUI/NRAO and NAOJ.
\end{acknowledgements}

\bibliographystyle{aa}
\bibliography{aanda.bib}

\end{document}